\documentclass[11pt,letterpaper]{HiPCStyle}
\usepackage{times}
\usepackage{epsfig}
\usepackage{amssymb}
\usepackage{amsmath}
\usepackage{amsfonts}
\usepackage{graphicx}
\usepackage{subfigure}
\usepackage{verbatim}
\usepackage{setspace}

\newcommand{\secref}[1]{Section~\ref{#1}}
\newcommand{\reshape}{ReSHAPE}

\makeatletter
\def\bstctlcite#1{\@bsphack
\@for\@citeb:=#1\do{%
\edef\@citeb{\expandafter\@firstofone\@citeb}%
\if@filesw\immediate\write\@auxout{\string\citation{\@citeb}}\fi}%
\@esphack}
\makeatother

\title{\bf ReSHAPE: A Framework for Dynamic Resizing and Scheduling of Homogeneous Applications in a  Parallel Environment\\ 
}

\author{Rajesh Sudarsan and Calvin J. Ribbens\\
Department of Computer Science\\
Virginia Tech, Blacksburg, VA 24061-0106\\
\texttt {\{sudarsar, ribbens\}@vt.edu}\\
       }
\date{}

\begin{document}

\maketitle
\thispagestyle{empty}

\begin{abstract}
\singlespace
Applications in science and engineering often require huge computational resources for solving problems within a reasonable time frame. Parallel supercomputers provide the computational infrastructure for solving such problems. A traditional application scheduler running on a parallel cluster only supports static scheduling where the number of processors allocated to an application remains fixed throughout the lifetime of execution of the job. Due to the unpredictability in job arrival times and varying resource requirements, static scheduling can result in idle system resources thereby decreasing the overall system throughput. In this paper we present a prototype framework called \reshape, which supports dynamic resizing of parallel MPI applications executed on distributed memory platforms. The framework includes a scheduler that supports resizing of applications, an API to enable applications to interact with the scheduler, and a library that makes resizing viable. Applications executed using the \reshape{} scheduler framework can expand to take advantage of additional free processors or can shrink to accommodate a high priority application, without getting suspended. In our research, we have mainly focused on structured applications that have two-dimensional data arrays distributed across a two-dimensional processor grid. The resize library includes algorithms for processor selection and processor mapping. Experimental results show that the \reshape{} framework can improve individual job turn-around time and overall system throughput.
\end{abstract}

\bstctlcite{IEEEexample:BSTcontrol}

\section{Introduction}
\label{sec:Introduction}

As terascale supercomputers become more common, and as the high-performance compu\-ting (HPC) community turns its attention to petascale machines, the challenge of providing effective resource management for high-end machines grows in both importance and difficulty. High capability computing resources are by definition expensive, so the cost of underutilization is high, e.g., wasting 5\% of the compute nodes on a 10,000 node cluster is a much more serious problem than on a 100 node cluster. Furthermore, high capability computing is characterized by long-running, high node-count jobs. A job-mix dominated by a relatively small number of such jobs is notoriously difficult to schedule effectively on a large-scale cluster. In over three years of experience operating System~X, a terascale system at Virginia Tech, we have observed that conventional schedulers are hard-pressed to achieve over 90\% utilization with typical job-mixes. Of course, near 100\% utilization can be achieved with a ready supply of embarrassingly-parallel short-duration or suspendable jobs; but our aim has been to move such work onto inexpensive Grid platforms, while reserving System~X for work requiring high-capability computing. A fundamental problem is that conventional parallel schedulers are static. That is, once a job is allocated a set of resources, it continues to use those same resources until it finishes execution.  It is worth asking whether a dynamic resource manager, which has the ability to modify resources allocated to jobs at runtime, would allow more effective resource management. The focus of our research is on dynamically reconfiguring parallel applications to use a different number of processes, i.e., on \textit{dynamic resizing} of applications.

It is not difficult to identify several potential benefits of dynamic resizing.  From the perspective of the scheduler, dynamic resizing would allow higher machine utilization and job throughput.  For example, under static scheduling it is common to see jobs stuck in the queue because they require just a few more processors than are currently available.  With resizing, the scheduler may be able to start (i.e., \textit{back-fill}) them earlier, squeezing them into processors that are available, and then adding more processors later.  Alternatively, the scheduler could add unused processors to a particular job in order to help that job finish earlier, thereby freeing up resources for waiting jobs. Schedulers could also expand or contract the processor allocation for an already running application in order to accommodate higher priority jobs, or to meet a quality of service (QoS) or advance reservation commitment. 

Dynamic resizing is also attractive from the individual job's perspective. Any scheduling mechanism that allows a job to start earlier or gain more processors later can improve the turn-around time for that job. Applications that consist of multiple phases, some of which are more computationally intense than others, could also benefit from resizing to the most appropriate node count for each phase.  Another interesting motivation is in identifying a node count \textit{sweet spot} for a particular job. For any parallel application, when the problem size is fixed, there is a point beyond which adding more processors helps little or not at all. Dynamic resizing gives applications and schedulers the opportunity to probe for good node-counts for a particular application and problem.

In order to explore the potential benefits and challenges of dynamic resizing, we are developing ReSHAPE, a framework for dynamic {\bf Re}sizing and {\bf S}cheduling of {\bf H}omogeneous {\bf A}pplications in a {\bf P}arallel {\bf E}nvironment. The ReSHAPE framework includes a programming model and API, data redistribution algorithms and a runtime library, and a parallel scheduling and resource management system.  
In order for \reshape{} to be a usable and effective framework,
there are several important criteria which these components should meet.
The programming model needs to be simple enough so that existing code can be ported to the new system without an unreasonable re-coding burden.  Runtime mechanisms must include support for releasing and acquiring processors and for efficiently redistributing application state to a new set of processors.  The scheduling mechanism must exploit resizability to increase system throughput and reduce job turn around time. In this paper we describe the initial design and implementation of ReSHAPE, along with experimental results which illustrate the potential benefits of resizing.
For the moment, we focus on improving the performance of iterative applications using \reshape{}. But the applicability of this framework can be extended to non-iterative applications as well. The main contributions of our framework include:
\begin{enumerate}
\item providing a robust framework to schedule and dynamically manage resources for parallel applications executing in a homogeneous cluster.
\item an efficient runtime library for processor remapping and 2-D data redistribution.
\item a simple programming model and easy-to-use API to port existing scientific applications to run with the ReSHAPE framework.
\end{enumerate}

The rest of the paper is organized as follows. \secref{sec:RelatedWork} presents  related work in the area of dynamic processor allocation. \secref{sec:SystemOrganization} describes the design features and implementation details of the \reshape{} framework. \secref{sec:Experimental} describes the experimental methodology used for evaluating the performance of the framework and presents the results. And finally, \secref{sec:Conclusion} concludes with a summary and describes future work.

\section{Related Work}
\label{sec:RelatedWork}

Dynamic scheduling of parallel applications has been an active area of research for several years. Much of the early work targets shared memory architectures although several recent efforts focus on grid environments. 
Feldmann et al.~\cite{feldmann94dynamic} propose an algorithm for dynamic 
scheduling on parallel machines under a PRAM programming model. 
McCann et al.~\cite{mccann} propose a dynamic processor allocation policy 
for shared memory multiprocessors and study
space-sharing vs.\ time-sharing in this context.
Julita et al.~\cite{julita} present a scheduling policy for 
shared memory systems that allocates processors based on the 
performance of the application. 

Moreira and Naik~\cite{moreira} propose a technique for dynamic resource 
management on distributed systems using a checkpointing framework called 
Distributed Resource Management Systems (DRMS). 
The framework supports jobs that can change their active number of tasks during program execution, map the new set of tasks to execution units, 
and redistribute data among the new set of tasks. DRMS does not make 
reconfiguration decisions based on application performance however, and it uses 
file-based checkpointing for data redistribution.  
A more recent work by Kale~\cite{kale} achieves reconfiguration of MPI-based 
message passing programs.  However, the reconfiguration is achieved using 
Adaptive MPI (AMPI), which in turn relies on 
Charm++~\cite{charm} for the processor virtualization layer, and
requires that the application be run with many more threads than processors. 

Weissman et al.~\cite{weissman} describe an application-aware job scheduler 
that dynamically controls resource allocation among concurrently executing jobs.
The scheduler implements policies for adding or removing resources from jobs
based on performance predictions from the 
Prophet system~\cite{weissman99prophet}.  
All processors send data to the root node for 
data redistribution.  The authors present simulated results based on 
supercomputer workload traces.
Cirne and Berman~\cite{francine02} use the term \textit{moldable} to describe 
jobs which can adapt to different processor sizes. 
In their work the application scheduler AppLeS selects the job 
with the least estimated turn-around time out of a set of moldable jobs, 
based on the current state of the parallel computer. 
Possible processor configurations are specified by the user, and the number
of processors assigned to a job does not change after job-initiation time.

Vadhiyar and Dongarra~\cite{vadhiyar03SRS,vadhiyar03performance} 
describe a user-level checkpointing 
framework called Stop Restart Software (SRS) for developing malleable and 
migratable applications for distributed and Grid computing systems. 
The framework implements a rescheduler which monitors application progress 
and can migrate the application to a better resource. 
Data redistribution is done via user-level file-based checkpointing.

El~Maghraoui et al.~\cite{ElMaghraoui} describe a framework to enhance the 
performance of MPI applications through process checkpointing, migration and
load balancing. The infrastructure uses a distributed middleware framework that 
supports resource-level and application-level profiling for load balancing.
The framework continuously evaluates application performance, discovers new 
resources, and migrates whole or parts of applications to better resources. 
Huedo et al.~\cite{huedo2004gridway} also describe a framework, called Gridway,
for adaptive execution of applications in Grids. 
Both of these frameworks target Grid environments 
and aim at improving the resources assigned to an application by 
replacement rather than increasing or decreasing the number of resources.

The \reshape\ framework described in this paper has several aspects that 
differentiate it from the above work. \reshape{} is designed for applications
running on distributed-memory clusters. 
Like~\cite{francine02,weissman}, applications must be moldable in order to 
take advantage of \reshape\; but in our case the user is not required
to specify the legal partition sizes ahead of time.
Instead, \reshape\ can dynamically calculate partition sizes based on 
the run-time performance of the application.
Our framework uses neither file-based checkpointing nor a single node 
for redistribution.   Instead, we use an efficient data redistribution
algorithm which remaps data on-the-fly using message-passing over
the high-performance cluster interconnect.
Finally, we evaluate our system using experimental data from a real
cluster, allowing us to investigate potential benefits both for 
individual job turn-around time and overall system utilization and
throughput.

\section{System Organization}
\label{sec:SystemOrganization}

\begin{figure}[ht]
\subfigure[]{
\includegraphics[scale=0.38]{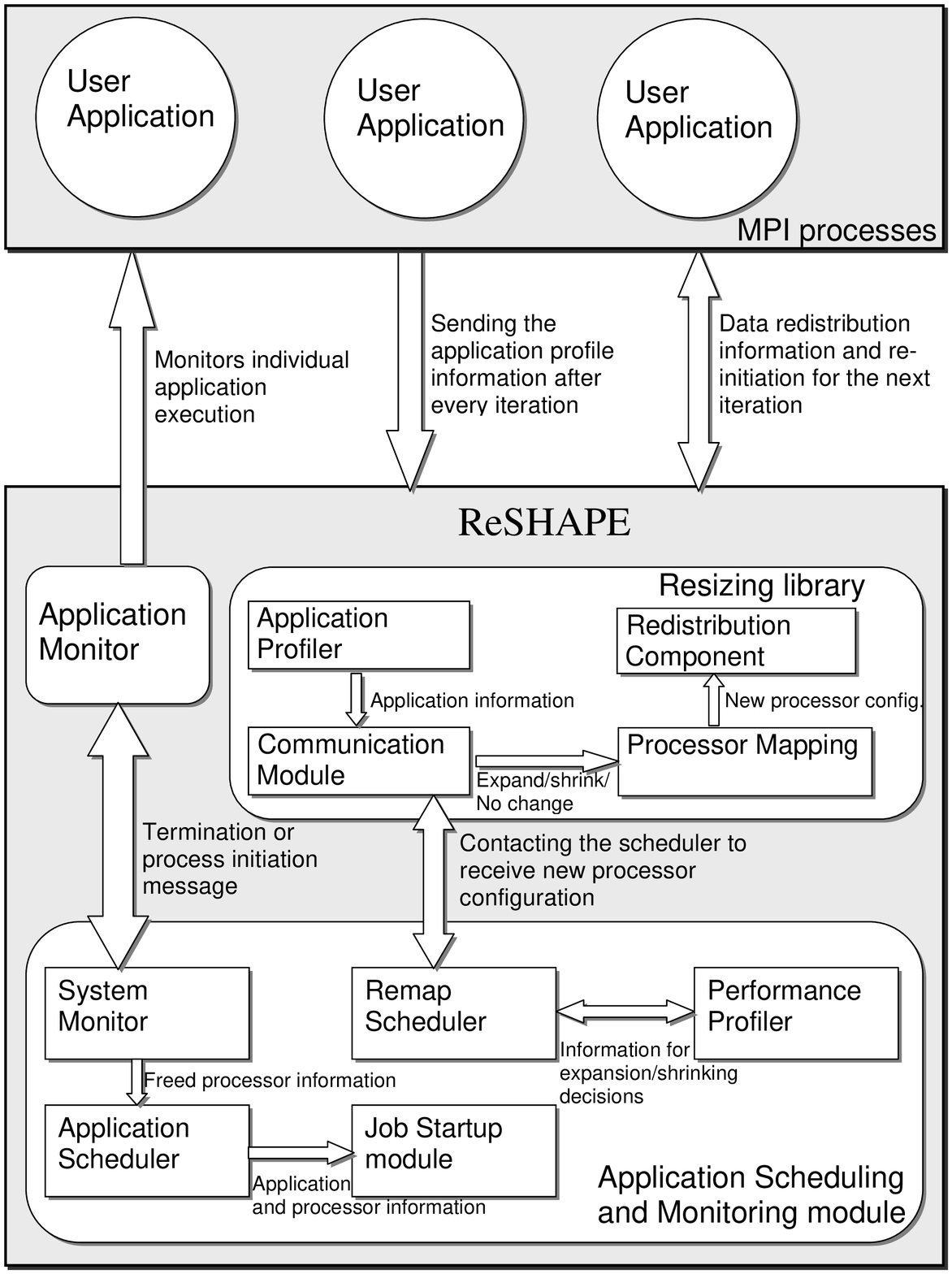}
\label{fig:design}
}
\subfigure[]{
\includegraphics[scale=0.38]{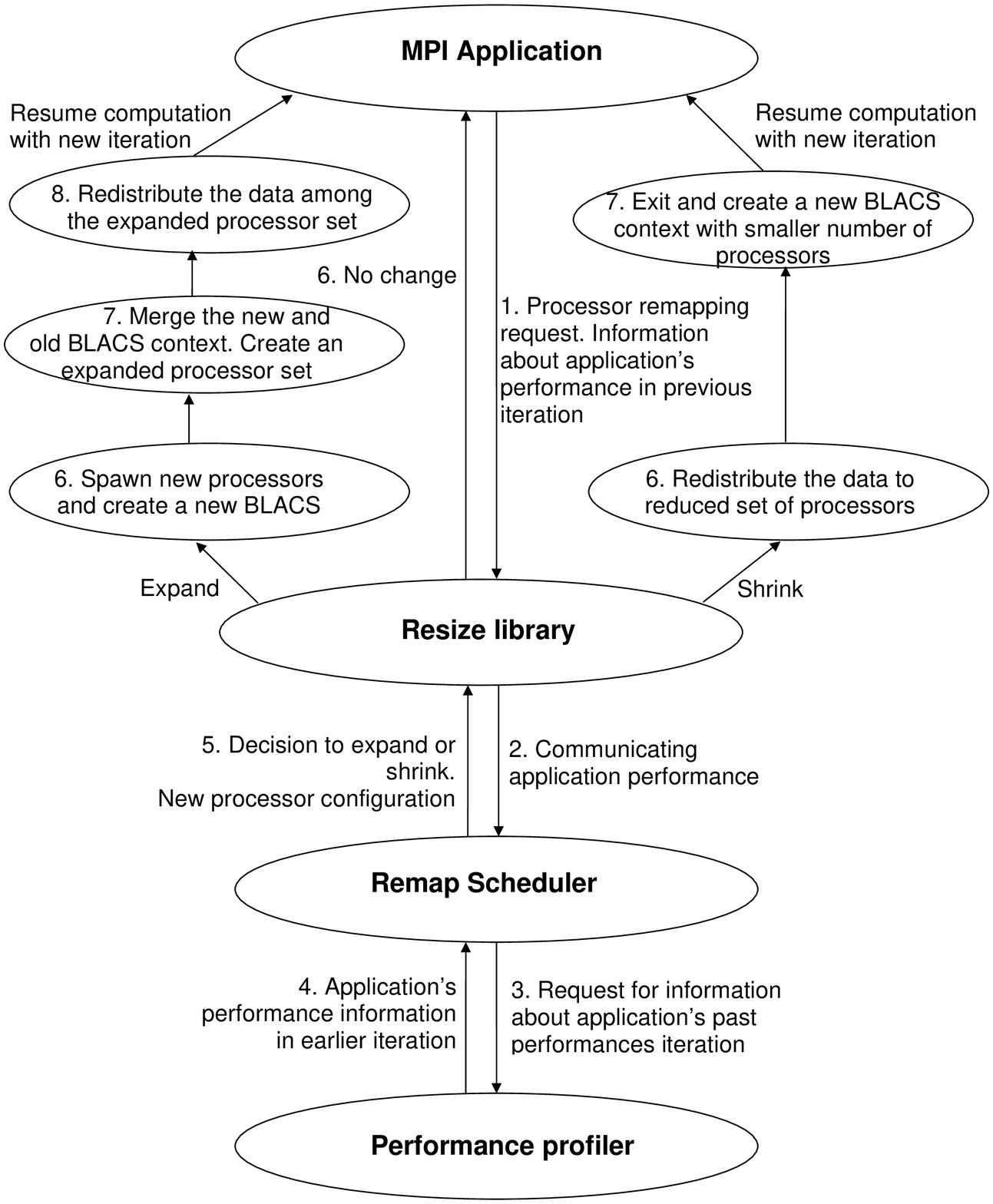}
\label{fig:statediagram}
}
\caption{(a) Architecture of \reshape. (b) State diagram for app\-lication ex\-pansion and shrinking }
\end{figure}

The \reshape{} framework, shown in Figure~\ref{fig:design}{}, consists of two main components. The first component is the application scheduling and monitoring module which schedules and monitors jobs and gathers performance data in order to make resizing decisions based on application performance, available system resources, resources allocated to other jobs in the system, and jobs waiting in the queue. The basic design of the application scheduler was originally part of the DQ/GEMS project~\cite{varadarajan}. 
The initial extension of DQ to support resizing was done by Chinnusamy and
Swaminathan~\cite{chinnuswamyMS04,swaminathanMS04}.
The second component of the framework consists of a programming model for resizing applications.  This includes a resizing library and an API for applications to communicate with the scheduler to send performance data and actuate resizing decisions. The resizing library includes algorithms for mapping processor topologies and redistributing data from one processor topology to another. 
 In our current implementation of the framework, the resizing library is built on top of the ScaLAPACK~\cite{scalapack} communication library, BLACS~\cite{dongarra}. 
We modified BLACS to support dynamic process management. It is important to note that the remapping algorithms are not specific to BLACS  and can be implemented as a generic resizing library. 
The resizing library currently includes generic one- and two-dimensional 
block-cyclic data redistribution algorithms for global arrays (see 
Section~\ref{subsec:Dataredist}), but it can be extended to support
other global data structures and other redistribution algorithms.

\reshape\ targets applications that are \textit{homogeneous} in two important ways. First, our approach is best suited to applications where data and computations are relatively uniformly distributed across processors. Second, we assume that the application is iterative, with the amount of computation done in each iteration being roughly the same.  While these assumptions do not hold for all large-scale applications,  they do hold for a significant number of 
large-scale scientific simulations. 
Hence, in the experiments described in this paper, we use applications with a single outer iteration, and with one or more large computations dominating each iteration.  Our API gives programmers a simple way to indicate \textit{resize points} in the application, typically at the end of each iteration of the outer loop. At resize points, the application contacts the scheduler and provides performance data to the scheduler.  The metric used to measure performance is the time taken to compute each iteration. 

\subsection{Application scheduling and monitoring module}
The application scheduling and monitoring module includes five components, 
each executed by a separate thread. The different components are System Monitor, Application Scheduler, Job Startup, Remap Scheduler, and Performance Profiler.
We describe each component in turn, along with a particular set of policies
that are implemented in our current system.  It is important to note that
the \reshape{} framework can easily be extended to support more 
sophisticated policies.

\noindent
\textbf{System Monitor.} An application monitor is instantiated on every compute node to monitor the status of an application executing on the node and report the status back to the System Monitor. If an application fails due to an internal error or finishes its execution successfully, the application monitor sends a job error or a job end signal to the System Monitor. The System Monitor then deletes the job and recovers the application's resources. For each application, only the monitor running on the first node of its processor set communicates with the System Monitor.

\noindent
\textbf{Application Scheduler.} An application is submitted to the scheduler for execution using a command line submission process. The scheduler enqueues the job and waits till the requested number of processors become available for execution. As soon as the resources become available, the scheduler selects the compute nodes, marks them as unavailable in the resource pool, and sends a signal to the job startup thread to begin  execution.
Our current implementation supports two basic resource allocation policies, First Come First Served (FCFS) and simple backfill.

\noindent
\textbf{Job Startup.} Once the Scheduler allocates the requested number of processors to a job, the job startup thread initiates an application startup process on the set of processors assigned to the job. The startup thread sends  job start information to the application monitor executing  on the first node of the allocated set. The application monitor sends job error or job completion messages back to the System Monitor.

\noindent
\textbf{Performance Profiler.} At every resize point, the Remap Scheduler receives performance data from a resizable application.
The performance data includes the number of processors used, iteration time taken to complete the previous iteration, and the redistribution time for mapping the global data from one processor set to another, if any.
The Performance Profiler maintains lists of the various processor sizes each 
application has run on and the performance of the application at each of those sizes.
The Profiler also maintains a list of possible shrink points of various applications and the anticipated impact on the application's performance. Each entry in the shrink points list consists of  job id of an application, number of processors each application can relinquish and the expected performance degradation which would result from this change. 
Note that applications can only shrink to processor configurations 
on which they have previously run.

\noindent
\textbf{Remap Scheduler.} The point between two subsequent iterations in an application is called a resize point. After each iteration, a resizable application contacts the Remap Scheduler with its latest iteration time.  The Remap Scheduler contacts the Performance Profiler to retrieve information about the application's past performance and  decides to expand or shrink the number of processors assigned to an application, as follows.  A decision to increase the number of processors given to an application is made if
    \begin{enumerate}
      \item there are idle processors in the system, and
      \item there are no jobs waiting to be scheduled on the idle processors, and
      \item there has been an improvement in the iteration time due to a previous expansion or the job has never been expanded to a bigger processor set.
    \end{enumerate}

The size and topology of the expanded processor set can be problem and
application dependent.  In our current implementation we
require that the global data be equally distributable across the new
processor set.  Furthermore, at job submission time applications can
indicate (in a simple configuration file) if they prefer 
a particular processor topology, e.g., a rectangular processor grid.
In the case where applications prefer ``nearly-square'' topologies, 
additional processors are added to the smallest row or column of the
existing topology.

The Remap Scheduler decides to shrink the number of processors for an 
application if it has previously run on a smaller processor set and
      \begin{enumerate}
        \item at the last resize point, the application expanded its processor set to a size that did not provide any performance benefit, or
        \item there are applications in the queue waiting to be scheduled. 
  \end{enumerate}

In the current implementation of \reshape, if there are jobs in the queue waiting to be scheduled, the Remap Scheduler 
checks whether the current application can give up a sufficient number of processors to schedule the next queued job. 
If it can, the application is resized to a smaller processor set, and the
freed processors allocated to the first job waiting in the queue.
If the application cannot free sufficient processors, the Remap Scheduler 
will shrink the application to its smallest shrink point 
(i.e., its starting processor set) and wait for the next application to 
check-in at its resize point. 
We emphasize again that more sophisticated policies are possible; indeed,
a significant motivation for \reshape{} in general, and the 
Performance Profiler in particular, is to serve as a platform for
research into more sophisticated resizing strategies.

\subsection {Resizing library and API}
The resizing library includes routines for changing the size of the processor set assigned to an application and for mapping processors and data from one processor set to another. An application needs to be  re-compiled with the resize library to enable the scheduler to dynamically add or remove processors to/from the application. During resizing, rather than suspending the job, the application execution control is transferred to the resize library which maps the new set of processors to the application and  redistributes the data (if any). Once mapping is completed, the resize library returns control back to the application and the application continues with its next iteration. The application user needs to indicate the global data structures and variables so that they can be redistributed to the new processor set after resizing. Figure~\ref{fig:statediagram}{} shows the different stages of execution required for changing the size of the processor set for an application.

\subsubsection{Processor remapping}
At resize points, if an application is allowed to expand to more processors, the response from the Remap Scheduler includes the size and the list of processors to which an application should expand. The resizing library spawns new processes on these  processors using the MPI-2 function MPI\_Comm\_spawn\_multiple called from the modified BLACS code. The inter-communicator between the spawned child processes and the parent processes are merged together to form a single intra-communicator for the expanded processor set. The old BLACS context is exited and a new context is created for the new processor set. A call to the redistribution routine remaps the global data to the new processor set. If the Scheduler asks an application to shrink, then the  application first redistributes its global data to the smaller processor subset, retrieves the previously stored MPI communicator for the application,  and creates a new BLACS context for the smaller processor set. The additional processes are terminated when the old BLACS context is exited.  The resizing library notifies the Remap Scheduler about the number of nodes relinquished by the application.

\subsubsection {Data redistribution}
\label{subsec:Dataredist}

The data redistribution library in ReSHAPE uses an efficient algorithm for redistributing block-cyclic  arrays between processor sets organized in a 1-D (row or column format) or checkerboard processor topology. The algorithm for redistributing the data among the checkerboard processor topology is an extension of the algorithm for a 1-D processor topology, initially proposed by Park et al.~\cite{park}. The algorithm uses a table based framework for index computation and a generalized circulant matrix formalism to compute an efficient contention-free communication schedule. The algorithm represents an initial (existing data layout) and a final data layout in a table format where the columns represent the old and resized processor sets respectively. A third table---called the destination processor table---contains the computed communication schedule. This table provides a mapping between the initial and final data layout. Each row in these tables represents a single communication step in which data blocks from the initial layout are transferred to their destination processors specified by their corresponding entries in the destination processor table. The data is transferred using MPI's persistent communication functions. 

\subsubsection {Application Programming Interface (API)}

A simple API allows user codes to access the \reshape{} framework and library. The core functionality is accessed through the following internal and external interfaces. 

\noindent
{\bf Advanced Functional API.} These functions are available for use by advanced application programmers. These functions provide the main functionality of the resizing library by contacting the scheduler, remapping the processors after an expansion or a shrink, and redistributing the data.
\begin {itemize}

\item    \textit{contact\_scheduler(node\_name, job\_id, num\_nodes, processor\_row\_count, \-processor\-\_column\-\_count, iteration\_time, redistribution\_time)}:
 contacts the scheduler, supplies last iteration time and redistribution time; on  return, the scheduler indicates whether the application should expand, shrink, or continue execution with the current processor size.

\item    \textit{expand\_processors()}: adds the new set of processors (defined by previous call to \-contact\_\-scheduler) to the current set using BLACS.

\item    \textit{shrink\_processors()}: reduces the processor set size (defined by previous call to \-contact\_\-scheduler) to an earlier configuration and relinquishes additional processors.

\item   \textit{Redistribute (Global data array, current BLACS context, new processor set size)}:  redistributes global data among the newly spawned processors. The redistribution time is computed and stored for next resize point.

\end{itemize}

\noindent
\textbf{Simple Functional API.} These functions are available for use by 
naive application programmers. In most cases, this simple API is sufficient 
for an application code to avail itself of the resize functionality of \reshape.

\begin{itemize}

\item \textit{log (iteration time)}: reads the individual iteration time from all the processes and logs the average iteration time in a file.

\item \textit{Resize()}: provides an abstraction layer and internally contacts the scheduler, reads the received information, expands or shrinks the processor set, and redistributes the data.
\end{itemize} 

The only other modifications needed to use \reshape{} with an
existing code involve code-specific local initialization of state
that may be needed when a new process is spawned and joins an already
running application.

\section{Experimental Results}
\label{sec:Experimental}

This section presents experimental results which demonstrate the potential of dynamic resizing for parallel applications.
The experiments were conducted on 50 nodes of a large  homogeneous cluster (System X).  Each node has two 2.3 GHz PowerPC 970 processors and 4GB of main memory.
Message passing was done using MPICH2~\cite{mpich2} over a Gigabit Ethernet
interconnection network.
We present results from two sets of experiments. The first set focuses
on the benefits of resizing for individual parallel applications; the
second set looks at improvements in cluster utilization and throughput
which result when several resizable applications are running concurrently.
We used five different applications for the experiments
(see Table~\ref{table:appdesc}).  In all cases, a single job consisted
of ten iterations of the task listed in the table, e.g., ten LU factorizations.
Table~\ref{table:procconfig} shows all the possible processor configurations used by the applications in the experimental setup.
LU and MM prefer a rectangular grid processor topology (as close to square
as possible), while Jacobi, Master-worker and FFT do not have any such requirement.
Our current implementation also requires that the number of processors
(in each dimension in the case of rectangular topologies) evenly divides
the problem size.

\begin{table}[!h]
\begin{center}
\caption{Workload application description }
\vspace{0.08in}
\footnotesize
\label{table:appdesc}
\begin{tabular}{|p{0.9in}|p{4.8in}|}
\hline
Application&Description\\
\hline
LU& LU factorization (PDGETRF from ScaLAPACK)\\
\hline
MM& Matrix-matrix multiplication (PDGEMM from PBLAS).\\
\hline
Master-worker & A synthetic master-worker application. Each iteration requires
20000 fixed-time work units.\\
\hline
Jacobi& An iterative jacobi solver (dense-matrix) application.\\
\hline
FFT& A 2D fast fourier transform application used for image transformation. \\
\hline
\end{tabular}
\end{center}
\end{table}

\begin{table}[!ht]
\begin{center}
\caption{Processor configuration for various problem sizes}
\vspace{0.08in}
\footnotesize
\label{table:procconfig}
\begin{tabular}{|p{1.2in}|p{4.5in}|}
\hline
Problem size&Processor configurations \\
\hline
8000 (LU, MM)& $1\times2$, $2\times2$, $2\times4$, $4\times4$, $4\times5$, $5\times5$, $5\times8$ \\
\hline
12000 (LU, MM)&$1\times2$, $2\times2$, $2\times3$, $3\times3$, $3\times4$, $4\times4$, $4\times5$, $5\times5$, $5\times6$, $6\times6$, $6\times8$ \\
\hline
14000 (LU, MM)&$2\times2$, $2\times4$, $4\times4$, $4\times5$, $5\times5$, $5\times7$, $7\times7$ \\
\hline
16000 (LU, MM)&$2\times2$, $2\times4$, $4\times4$, $4\times5$, $5\times5$, $5\times8$ \\
\hline
20000 (LU, MM)&$2\times2$, $2\times4$, $4\times4$, $4\times5$, $5\times5$, $5\times8$ \\
\hline
21000 (LU, MM)& $2\times2$, $2\times3$, $3\times3$, $3\times4$, $4\times5$, $5\times5$, $5\times6$, $6\times6$, $6\times7$, $7\times7$ \\
\hline
24000 (LU, MM)&$2\times4$, $3\times4$, $4\times4$, $4\times5$, $5\times5$, $5\times6$, $6\times6$, $6\times8$ \\
\hline
8000~(Jacobi)&$4, 8, 10, 16, 20, 32, 40,50$ \\
\hline
8192~(FFT)&$2, 4, 8, 16, 32$ \\
\hline
20000~(Master-worker)&$4, 6, 8, 10, 12, 14, 16, 18, 20, 22$ \\
\hline
\end{tabular}
\end{center}
\end{table} 

\subsection{Performance benefits for individual applications}
Although in practice applications rarely have an entire cluster
to themselves, it is not uncommon to have a number of
available processors for part of the running-time of large
applications.  We can use those available processors to
increase the performance of a given application, or perhaps
to discover that additional resources are not beneficial
for a given application.
The \reshape\ Performance Profiler and Remap Scheduler uses
this technique to probe for ``sweet spots" in processor allocations
for a given application and to monitor the tradeoff between improvements in
performance and the overhead of data redistribution.

\subsubsection{Adaptive sweet spot detection}
\label{adaptivesweetspot}
An appropriate definition of sweet spot for a parallel application depends
on the context.  An application-centric view
is that the sweet spot is the point at
which adding processors no longer helps reduce execution time.
A more system-centric view would look at relative speedups when 
processors are added, and at the requirements and potential speedups of
other applications currently running on the system.
With the performance data gathering and resizing capabilities of \reshape{} 
we are exploring different approaches to sweet spot definition and detection.

\begin{figure}[ht]
\subfigure[]{
\includegraphics[scale=0.63]{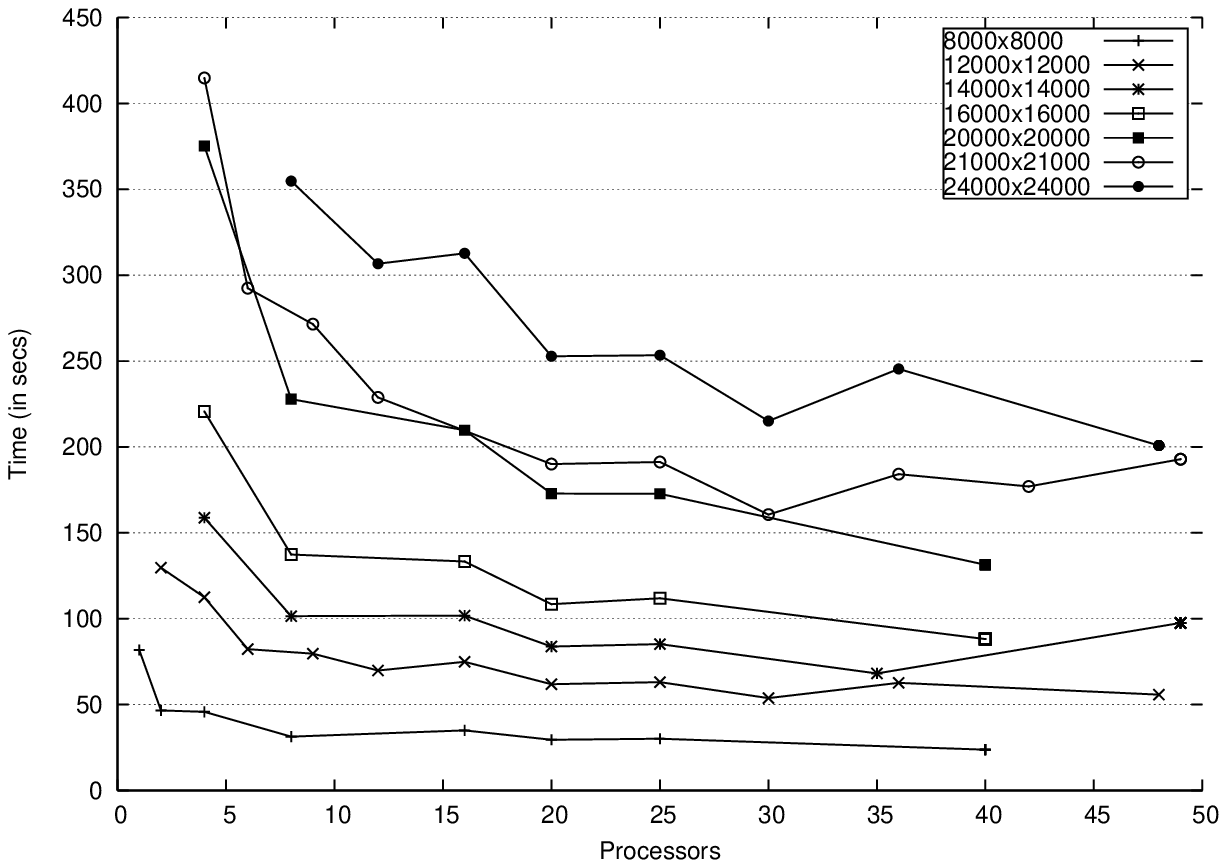}
\label{fig:sweetspot}
}
\subfigure[]{
\includegraphics[scale=0.63]{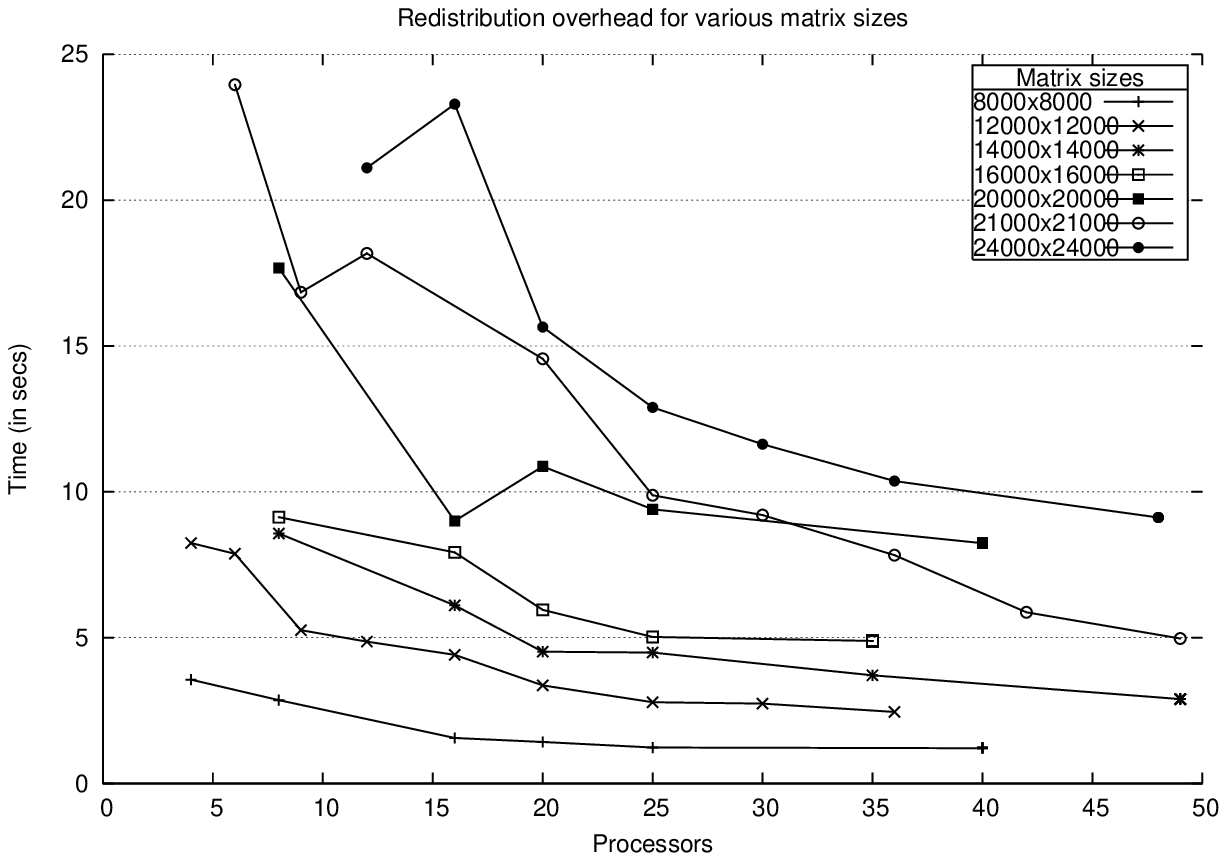}
\label{fig:redistribution}
}
\caption{(a) Running time for LU factorization with various matrix sizes. (b) Redistribution overhead for expansion for different matrix sizes.}
\end{figure}

To get an idea of the potential of sweet spot detection, consider the data
in Figure~\ref{fig:sweetspot} which shows the performance of LU 
for various problem sizes and processor configurations.
We observe that the performance benefit of resizing for
smaller matrix sizes is not high.  In this case, ReSHAPE can easily detect
that improvements are vanishingly small after the first few processor
size increases.
As expected, the performance benefits are greater for larger problem sizes.
For example, in the case of a $24000 \times 24000$ matrix, the iteration time
improves by 19.1\% when the processor set increases from 16 to 20 processors.
Our initial implementation of sweet spot detection in \reshape{} simply adds 
processors as long as they are available and as long as there is improvement
in iteration time.  If an application grows to a configuration that
yields no improvement, it is shrunk back to its most recent configuration.
This strategy yields reasonable sweet spots in all cases.  For example, problem 
size 21000 has a sweet spot at 30 processors.
However, the figure suggests a more sophisticated sweet spot
detection algorithm (under development) which uses performance over several 
configurations to detect relative improvements below some required threshold, or when
resources are available probes configurations beyond the largest 
considered so far, to see if performance depends strongly on particular
processor counts.

\subsubsection{Redistribution overhead}

Every time an application adds or releases processors, the globally
distributed data has to be redistributed to the new processor topology.
Thus, the application incurs a redistribution overhead each time it expands or shrinks.
As an example,
Figure~\ref{fig:redistribution} shows the overhead for
redistributing large dense matrices
for different matrix sizes using the \reshape\ resizing library.
Each data point in the graph represents the data redistribution cost
incurred when increasing the size of the processor configuration from the
previous (smaller) configuration.
Problem size 8000 and 12000 start execution with 2 processors,
problem size 14000, 16000, 20000 and 21000 start with 4 processors, and the 24000
case starts with 8 processors.
The starting processor size is the smallest size which can accommodate the data. The trend shows that the redistribution cost increases with matrix size,
but for a particular matrix size
the cost decreases as we increase the number of processors. This makes sense because for small processor size, the amount of data per processor that must be transferred is large.
By tracking the data redistribution costs as applications are resized,
\reshape\ can better weigh the potential benefits of resizing a
particular application.

\begin{figure}[ht]
\subfigure[]{
\includegraphics[scale=0.35]{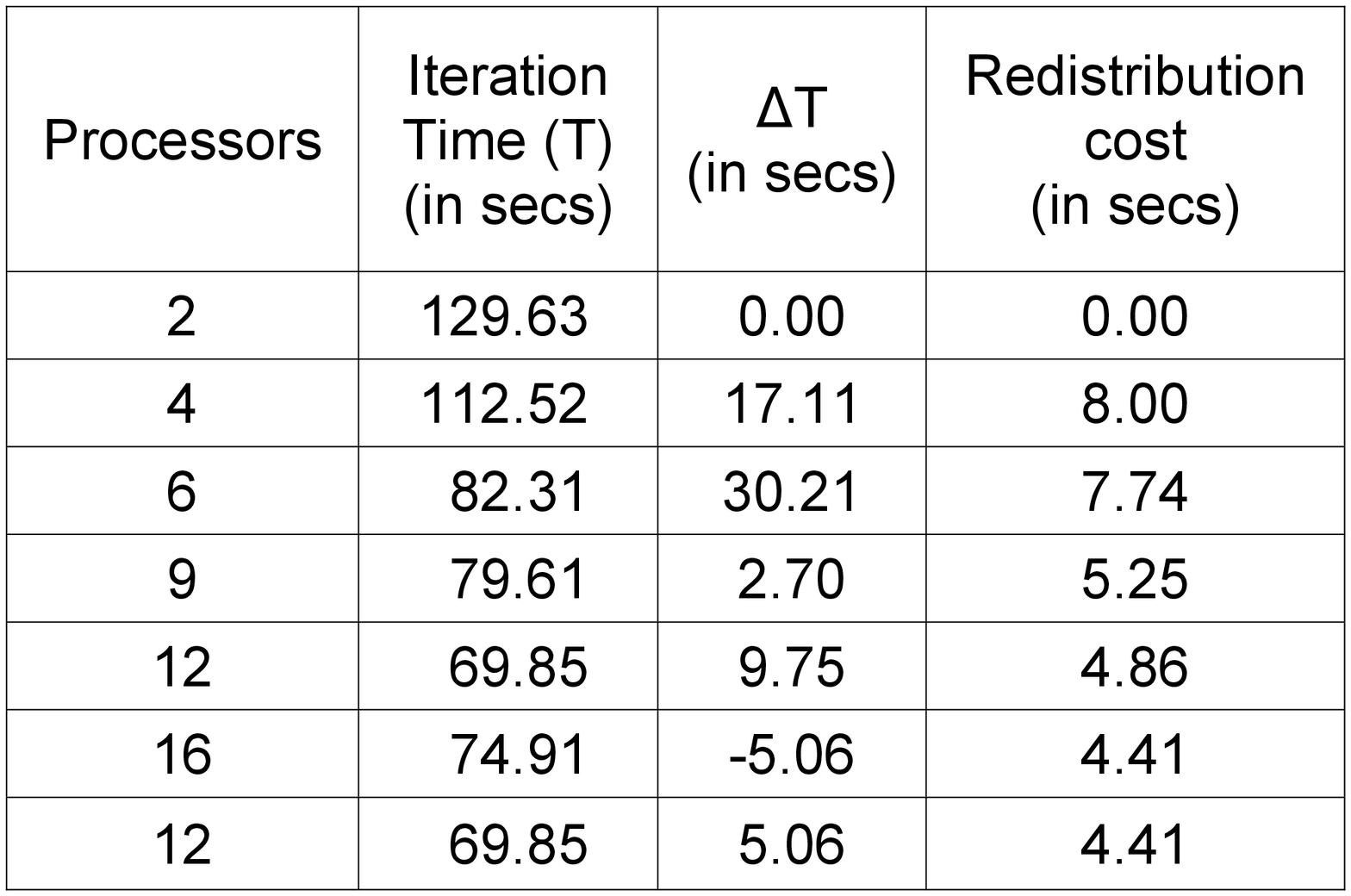}
\label{fig:overhead}
}
\subfigure[]{
\includegraphics[scale=0.7]{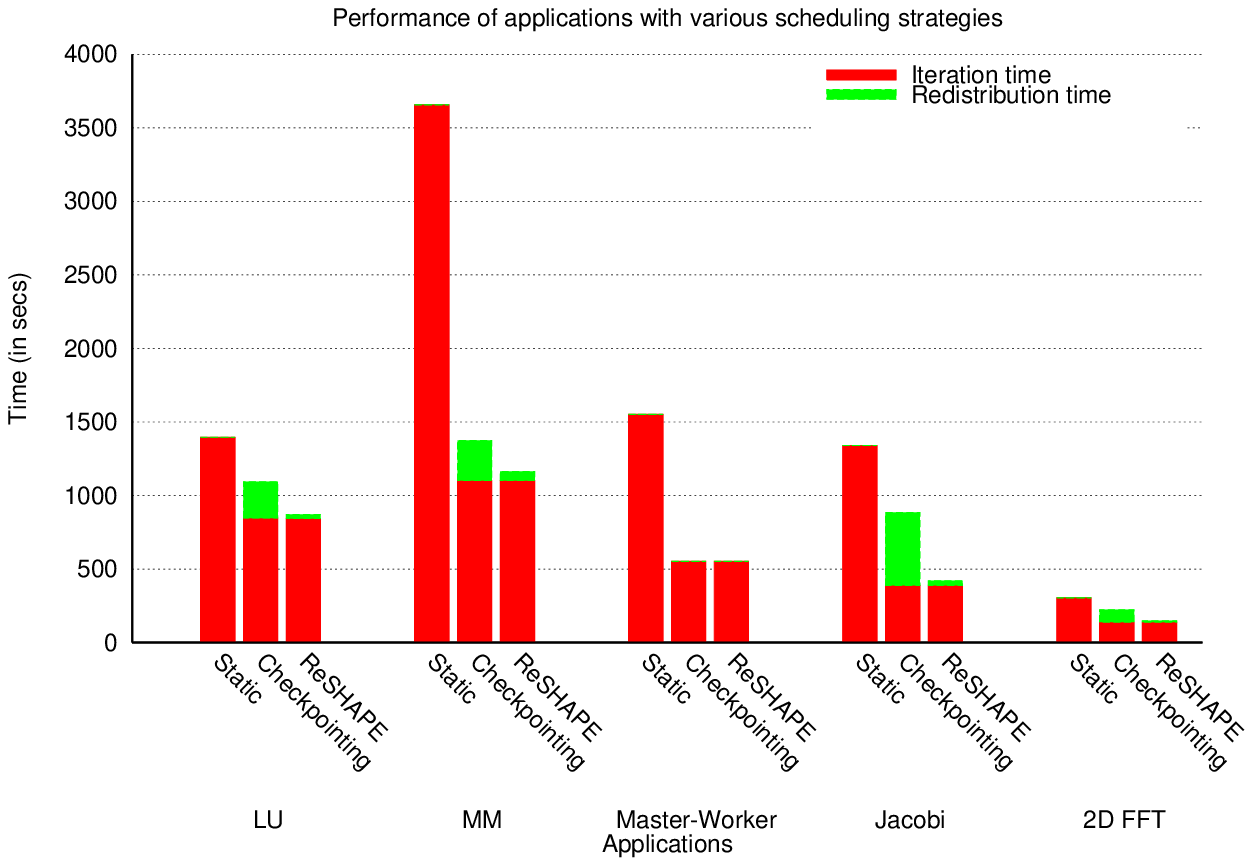}
\label{fig:overallcharac}
}
\caption{(a) Iteration and redistribution for LU on problem size 12000. 
(b) Performance with static scheduling, dynamic scheduling with file-based checkpointing, and dynamic scheduling with ReSHAPE redistribution, for 
LU(problem size 1200), MM(14000), Master-worker, Jacobi(8000), FFT (8192).
LU, MM, Jacobi and Master-worker start with 4 processors. FFT starts with 2 processors.}

\end{figure}

Figure~\ref{fig:overhead} compares the improvement in iteration time with
overhead cost involved in redistribution for the $12000\times12000$ matrix.  The LU application is executed for 10 iterations and the timing results are recorded after every iteration.
(Iterations 8-10 are not shown since \reshape{} held the processor allocation at 12.)
The application starts execution with 2 processors and gradually increases the processor size, maintaining its nearly-square topology. 
In this case, the cost of redistribution from 2 to 4 processors and from 4 to 6 processors is compensated by improvement in time in a single iteration. 
But when the application expands from 6 to 9 processors, more than one iteration is necessary to compensate for the overhead. 
This situation is more common in applications that are more communication intensive than computation intensive. 
In such applications, it is better to execute many iterations before a single redistribution occurs. 
A complex prediction strategy can estimate the redistribution cost~\cite{wolski1997pcr} for an application for a particular processor size. 
The remap scheduler could use this estimate to make resizing decisions. 
However, with \reshape{} we save a record of actual redistribution costs
between various processor configurations, which allows for more informed
decisions.
In Figure~\ref{fig:overhead},  $\Delta T = -5.06$ indicates that the 
performance degraded when the 
processor size expanded from 12 to 16 processors.  Hence, the scheduler resized 
the application back to 12 processors, with an overhead of 4.41 seconds of 
redistribution time. 
We noted a similar behavior in the overhead cost and iteration timings with 8000, 16000, 2000, and 24000 problem sizes.

To get an idea of the relative overhead of redistribution using
the \reshape\ library compared to file-based checkpointing,
we implemented a simple checkpointing library in which 
all data is saved and restored through a single node.
Figure~\ref{fig:overallcharac} shows the computation (iteration) time and
redistribution time with 
data redistribution by file-based checkpoint/restart and by the
\reshape\ redistribution algorithm. 
In both cases dynamic resizing is provided by \reshape.
The time for static scheduling (i.e., keeping the processor allocation constant
at the initial value for the entire computation) is also shown in the figure.
The results are from 10 iterations of each application using the processor
configurations in Table~\ref{table:procconfig}, and with \reshape{} adding
processors until a sweet spot is detected.
Redistribution via checkpointing is obviously much more expensive than
using our message-passing based redistribution algorithm.  For example,
with LU, checkpointing is 8.3 times more expensive than \reshape\ redistribution.
For matrix multiplication, Jacobi, and 2D FFT, the relative cost of
checkpointing vs.\ \reshape\ redistribution is
4.5, 14.5, and 7.9, respectively.
The master-worker application has no data to redistribute and hence
shows no difference between checkpointing and \reshape.

\subsection{Performance benefits with multiple applications}

The second set of experiments involves concurrent scheduling of a variety of
jobs on the cluster using the \reshape{} framework. In this case, we can safely assume that when multiple jobs are concurrently scheduled in a system and are competing for resources, not all jobs can adaptively realize and run at their sweet spot for the entirety of their execution. A job might be forced to shrink even before reaching its sweet spot to accommodate new jobs. Since the applications that are closer to their sweet spot do not show high performance gains, they become the most probable candidates for shrinking decisions. The long running jobs are shrunk to a smaller size without significant performance hit thereby benefiting shorter jobs waiting in the queue.
The \reshape{} framework also benefits the long running jobs as they can utilize resources beyond their initial allocation. These scenarios are not possible in a conventional static scheduler where the jobs in the queue wait till the resources become available.

We evaluated \reshape{}'s performance using two different workloads.
Each workload constituted a unique combination of the five applications
described in Table~\ref{table:appdesc}.
Table~\ref{table:workload} shows the problem sizes used for each workload.

\begin{table}[!h]
\begin {center}
\caption{Workloads for job mix experiment  }
\vspace{0.08in}
\footnotesize
\label{table:workload}
\begin{tabular}{|l|l|}
\hline
Workload&Applications\\
\hline
W1&LU($21000\times21000$), MM($14000\times14000$), Master-worker($4000000000$), Jacobi($8000\times8000$),\\
&FFT ($8192\times8192$) \\
\hline
W2 &LU($21000\times21000$), Master-worker($4000000000$), Jacobi($8000\times8000$), FFT ($8192\times8192$)\\
\hline
\end{tabular}
\end{center}
\end{table}

\subsubsection{Workload 1}
\begin{figure}[ht]
\subfigure[]{
\includegraphics[scale=0.63]{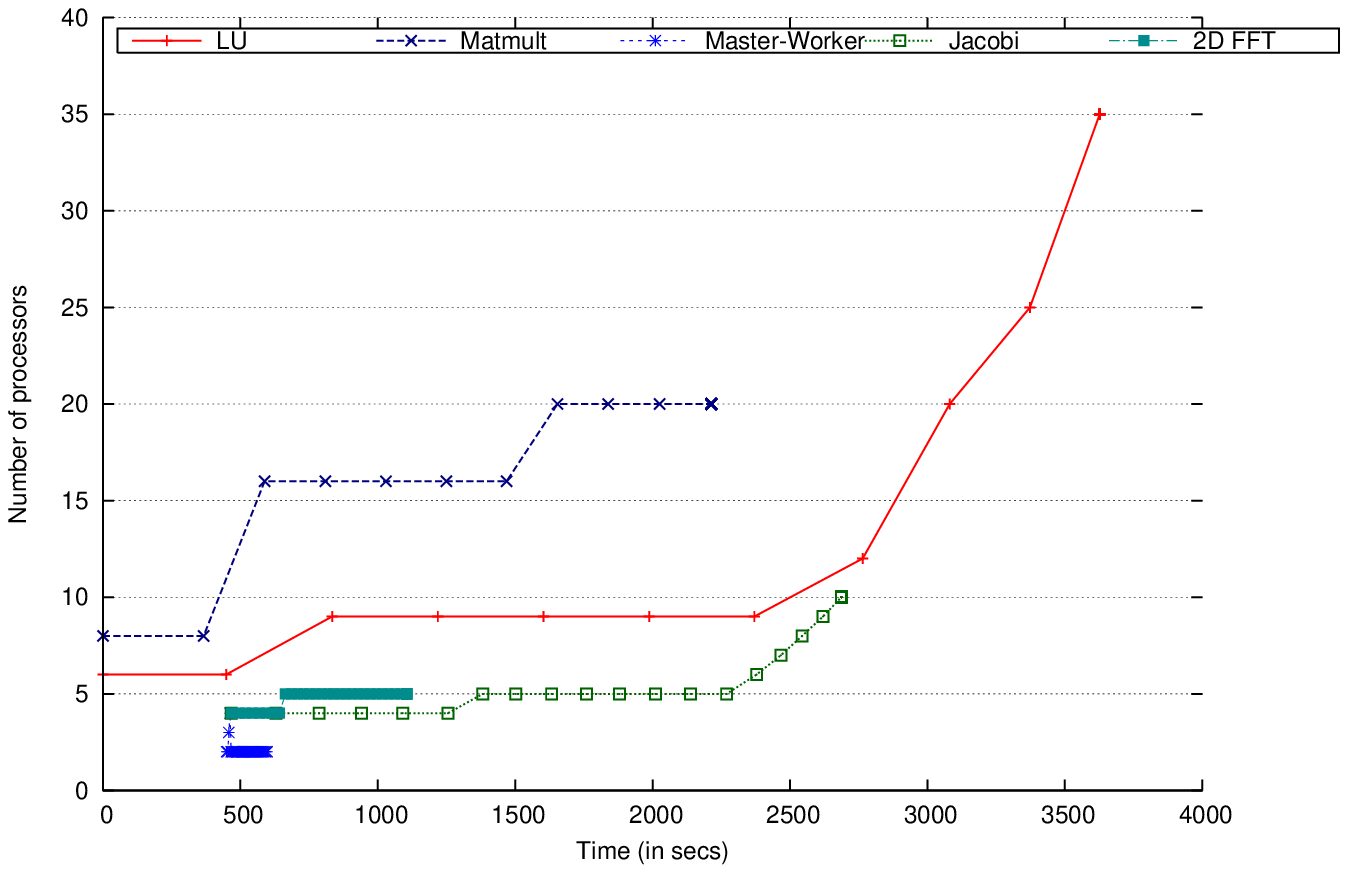}
\label{fig:jobmix1}
}
\subfigure[]{
\includegraphics[scale=0.63]{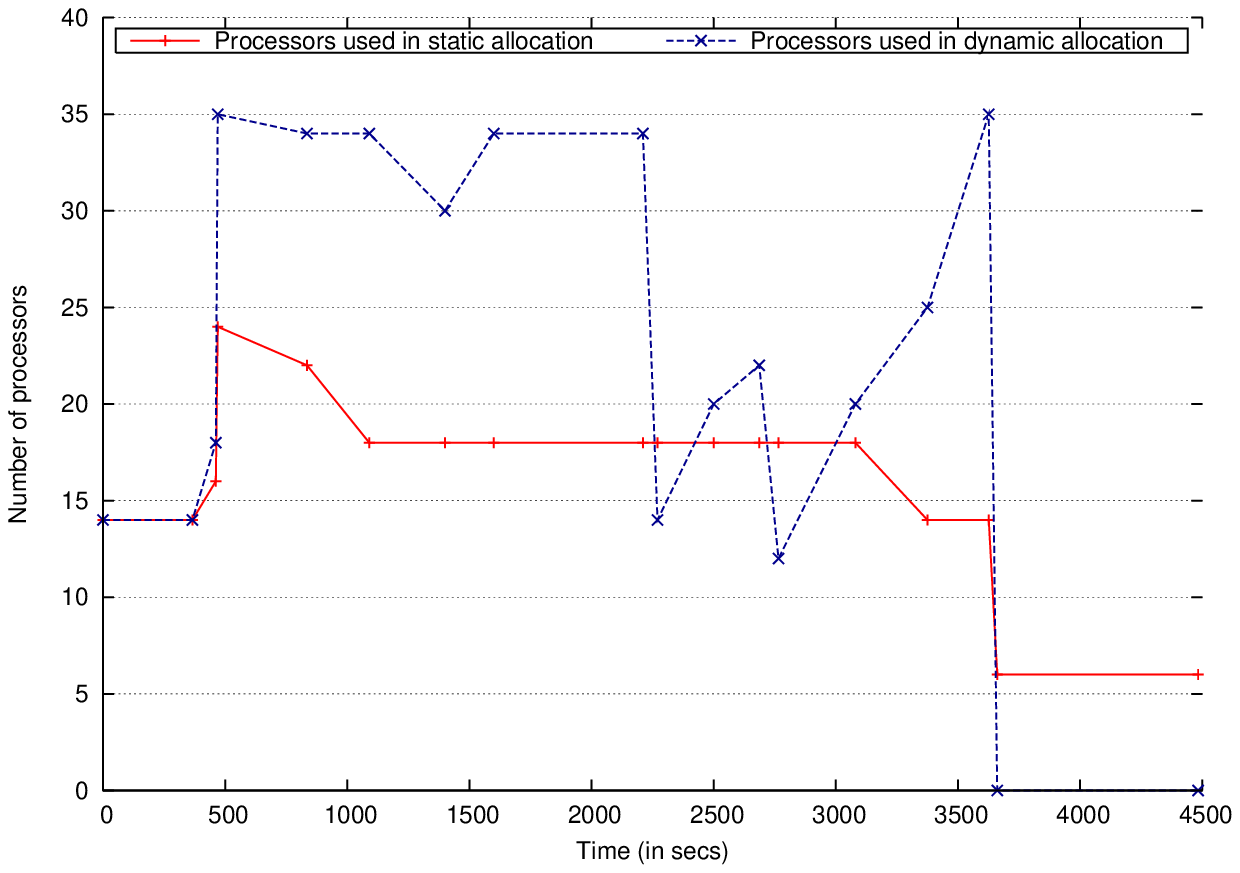}
\label{fig:processor_usage1}
}
\caption{(a) Processor allocation history for workload 1.  
(b) Total processors used for workload 1.}
\end{figure}

Figure~\ref{fig:jobmix1} shows the number of processors allocated to each job for the lifetime of the five jobs scheduled by the ReSHAPE framework. The total number of processors available is 36.  The data sizes considered for this experiment are limited by the availability of number of processors. On a larger cluster
comprising hundreds or even thousands of processors, much larger applications can execute---larger in terms of both data size and processor allocations.
The LU and the matrix multiplication application (MM) are scheduled at \textit{t=0} seconds. The Master-worker application arrives at \textit{t=450} seconds and is scheduled accordingly. The Jacobi and the FFT applications arrive at \textit{t=465} seconds. Since the master-worker application had the next earliest resize point, it shrunk by 2 processors to accommodate the newly queued jobs. As there were no other running or queued jobs in the system after \textit{t=2764} seconds, the LU application expanded to the maximum number of processors.
Figure~\ref{fig:processor_usage1} shows the total number of busy processors at
any time for both static and \reshape\ scheduling.

\begin{table}[ht]
\begin{center}
\caption{Job turn-around time}
\vspace{0.08in}
\footnotesize
\label{tab:jobmix1_throughput}
\begin{tabular}{|c|c|r@{\;\;\;\;\;\;\;\;\;\;}|r@{\;\;\;\;\;\;\;\;\;\;\;}|r@{\;\;\;\;\;}|}
\hline
Job&Initial proc. alloc.&\multicolumn{1}{|c|}{Static  scheduling}&\multicolumn{1}{|c|}{Dynamic scheduling}&\multicolumn{1}{|c|}{Difference}\\
&& \multicolumn{1}{|c|}{Time(in sec)}&\multicolumn{1}{|c|}{Time(in sec)}&\multicolumn{1}{|c|}{(in sec)}\\
\hline
LU &6&4482.60&3626.93&855.67\\
\hline
MM &8&3661.20&2212.72&1448.48\\
\hline
Master-worker&2&147.47&148.00&-0.53\\
\hline
Jacobi &4&3266.40&2220.17&1046.23\\
\hline
2D FFT &4&840.00&637.94&202.06\\
\hline
\end{tabular}
\end{center}
\end{table}

\newpage
Table~\ref{tab:jobmix1_throughput} shows the improved execution time for the
applications in workload 1. The average processor utilization\footnote{Utilization is defined as
the percentage of total available cpu-seconds that are assigned to a running
job.}
with static scheduling with workload 1 is 39.7\% whereas the average processor utilization using ReSHAPE dynamic scheduling is 70.7\%.
Note that Master-worker did not benefit from
resizing at all because it finished executing before additional processors
became available.

\subsubsection{Workload 2}

\begin{figure}[ht]
\subfigure[]{
\includegraphics[scale=0.63]{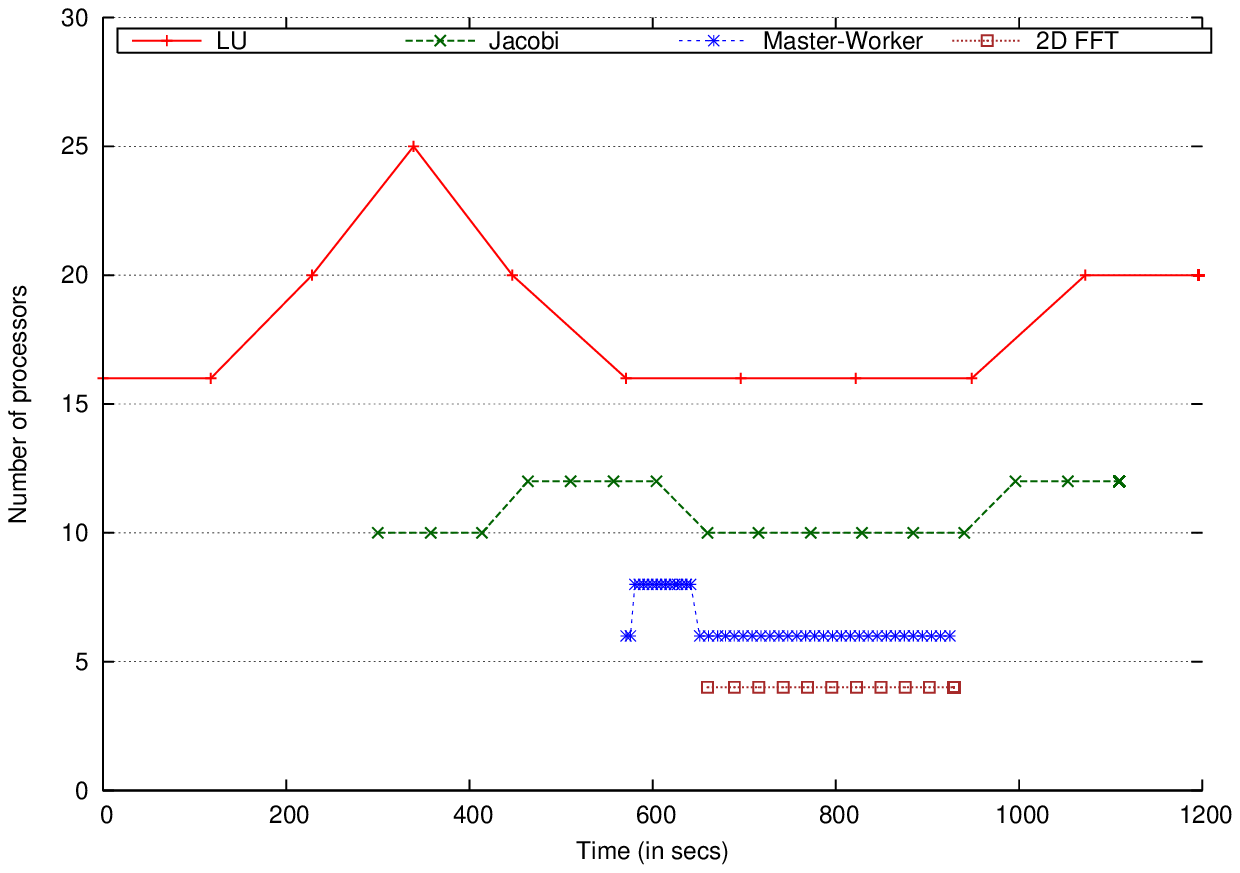}
\label{fig:jobmix2}
}
\subfigure[]{
\includegraphics[scale=0.63]{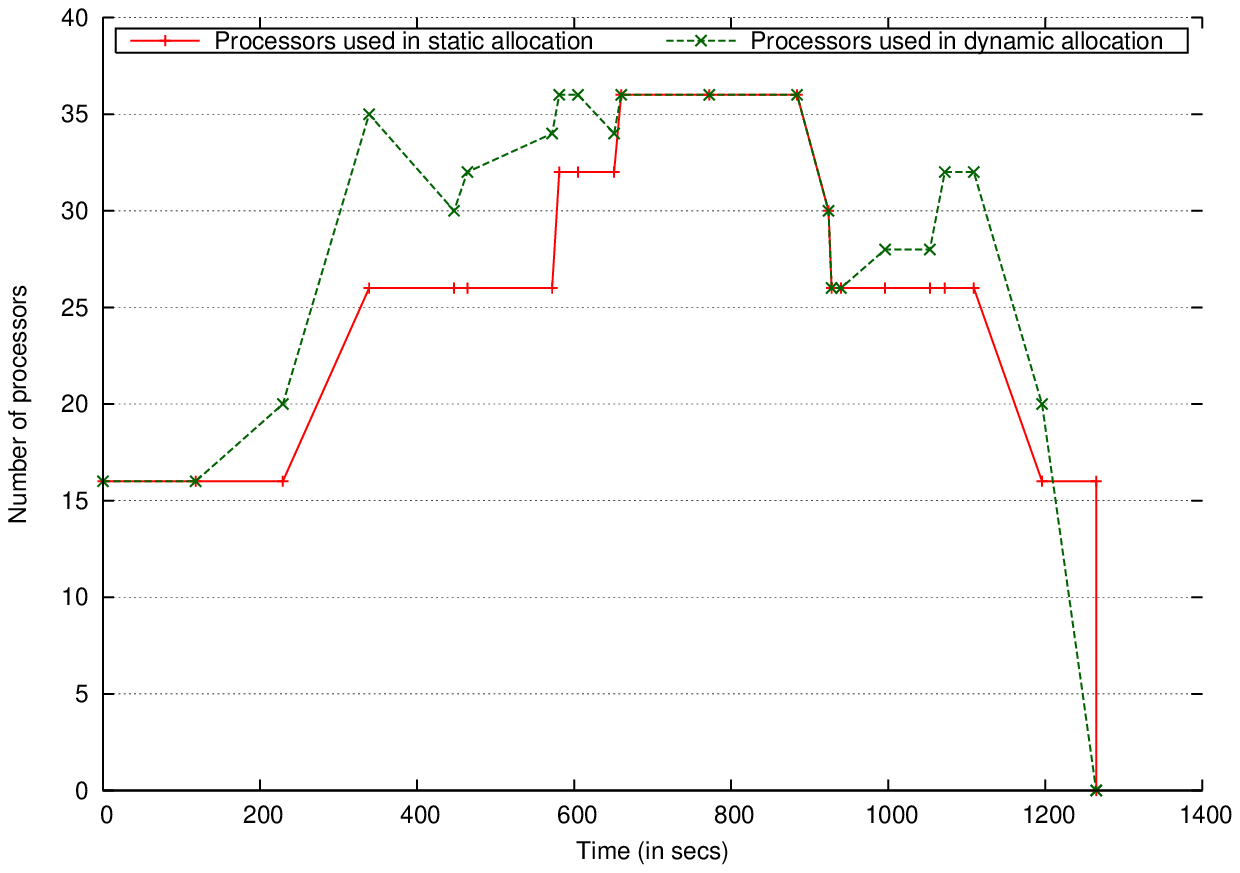}
\label{fig:processor_usage2}
}
\caption{(a) Processor allocation history for workload 2.  
(b) Total processors used for workload 2.}
\end{figure}

Workload 2 (see Figure~\ref{fig:jobmix2}) illustrates how an application
scheduled using the ReSHAPE framework shrinks to accommodate queued jobs.
The LU application expanded to 25 processors at \textit{t=338} seconds and located its sweet spot execution with 20 processors at \textit{t=446} seconds. 
The master-worker application that arrived at \textit{t=560} seconds was queued till \textit{t=571} seconds when the LU application shrunk to 16 processors to accommodate it. 
Similarly, the Master-worker application shrunk from 8 processors to 6 processors at \textit{t=660} seconds to accommodate the static scheduled 2D FFT application queued at \textit{t=650} seconds. 
Figure~\ref{fig:processor_usage2} shows that the dynamic scheduling has only a small advantage over static scheduling in the case of workload 2. The reason is due to the fact that the applications executed at their initial processor allocation for a greater part of their computation lifetime. Table~\ref{tab:jobmix2_throughput} compares the execution time between static and dynamic scheduling for workload 2.

\begin{table}[ht]
\begin{center}
\caption{Job turn-around time}
\vspace{0.08in}
\footnotesize
\label{tab:jobmix2_throughput}
\begin{tabular}{|c|c|r@{\;\;\;\;\;\;\;}|r@{\;\;\;\;\;\;\;\;\;\;\;}|r@{\;\;\;\;\;}|}
\hline
Job&Initial proc. alloc.&\multicolumn{1}{|c|}{Static  scheduling}&\multicolumn{1}{|c|}{Dynamic scheduling}&\multicolumn{1}{|c|}{Difference}\\
&& \multicolumn{1}{|c|}{Time(in sec)}& \multicolumn{1}{|c|}{Time(in sec)}&\multicolumn{1}{|c|}{(in sec)}\\
\hline
LU&16&1265.87\;&1196.00\;&69.87\;\\
\hline
Jacobi&10&867.00\;&809.25\;&57.75\;\\
\hline
Master-worker&6&355.00\;&353.33\;&1.67\;\\
\hline
2D FFT&4&268.69\;&268.69\;&0.00\;\\
\hline
\end{tabular}
\end{center}
\end{table}

\section{Conclusions and Future Work}
\label{sec:Conclusion}

In this paper we have introduced a framework that enables parallel message-passing
applications to be resized during execution.
The \reshape{} framework enables iterative applications to expand to
more processors, thereby automatically probing for potential performance
improvements.  When multiple applications are scheduled using \reshape, the system
uses performance results from the executing applications to select jobs
to shrink in order to accommodate new jobs waiting in the queue while
minimizing the negative impact on any one job.
These applications can later be expanded to larger processor sizes if the
system has idle processors.  The results show that applications executed
using \reshape{} demonstrate a significant improvement in job turn-around time
and overall system throughput.
The \reshape{} runtime library includes efficient algorithms for remapping
distributed arrays from one process configuration to another using
message-passing.  The system also records data redistribution times, so that
the overhead of a given resizing can be compared with the potential benefits
for long-running applications.
The \reshape{} API enables conventional SPMD (Single Program Multiple Data) programs
to be ported easily to take advantage of dynamic resizing.

We are currently working to take advantage of the \reshape{} framework to
evaluate scheduling strategies for processor reallocation, load-balancing,
quality-of-service and advanced reservation services.
Other current emphases include
adding resizing capabilities to several production scientific codes and
adding support for a wider array of distributed data structures and other data
redistribution algorithms.
Finally, we plan to make \reshape{} a more extensible framework so
that support for heterogeneous clusters, grid infrastructure, shared memory
architectures, and distributed memory architectures can be implemented  as
individual plug-ins to the framework.

\bibliographystyle{IEEEtranS}
\bibliography{reference}

\end{document}